\documentclass[%
 reprint,
preprintnumbers,
 amsmath,amssymb,
 aps,
prd,
 superscriptaddress,
]{revtex4-1}

\usepackage{siunitx}
\usepackage{graphicx}
\usepackage{dcolumn}
\usepackage{multirow}
\usepackage{bm}
\usepackage{mathtools}
\usepackage{lipsum}
\usepackage{hyperref}
\usepackage[mathlines]{lineno}
\begin{document}

\preprint{APS/123-QED}

\title{Design Studies Of A Pulsed Quasimonoenergetic 2-keV Neutron Source For Calibration Of Low Threshold Dark Matter Detectors}

\author{L. Chaplinsky} \affiliation{University of Massachusetts, Amherst Center for Fundamental Interactions and Department of Physics, Amherst, MA 01003-9337 USA}
\author{S. Fiorucci} \affiliation{Lawrence Berkeley National Laboratory, 1 Cyclotron Rd., Berkeley, CA 94720, USA}
\author{C.W. Fink} \affiliation{University of California Berkeley, Department of Physics, Berkeley, CA 94720, USA}
\author{M. Garcia-Sciveres} \affiliation{Lawrence Berkeley National Laboratory, 1 Cyclotron Rd., Berkeley, CA 94720, USA} \affiliation{International Center for Quantum-field Measurement Systems for Studies of the Universe and Particles (QUP,WPI), High Energy Accelerator Research Organization (KEK), Oho 1-1, Tsukuba, Ibaraki 305-0801, Japan}
\author{W. Guo} \affiliation{Department of Mechanical Engineering, FAMU-FSU College of Engineering, Florida State University, Tallahassee, FL 32310, USA} \affiliation{National High Magnetic Field Laboratory, Tallahassee, FL 32310, USA}
\author{S.A. Hertel} \affiliation{University of Massachusetts, Amherst Center for Fundamental Interactions and Department of Physics, Amherst, MA 01003-9337 USA}
\author{J.K. Wuko} \affiliation{University of Massachusetts, Amherst Center for Fundamental Interactions and Department of Physics, Amherst, MA 01003-9337 USA}
\author{X. Li} \affiliation{Lawrence Berkeley National Laboratory, 1 Cyclotron Rd., Berkeley, CA 94720, USA}
\author{J. Lin} \affiliation{University of California Berkeley, Department of Physics, Berkeley, CA 94720, USA}\affiliation{Lawrence Berkeley National Laboratory, 1 Cyclotron Rd., Berkeley, CA 94720, USA}
\author{R. Mahapatra} \affiliation{Texas A\&M University, Department of Physics and Astronomy, College Station, TX 77843-4242, USA}
\author{W. Matava} \affiliation{University of California Berkeley, Department of Physics, Berkeley, CA 94720, USA}
\author{D.N. McKinsey} \affiliation{University of California Berkeley, Department of Physics, Berkeley, CA 94720, USA} \affiliation{Lawrence Berkeley National Laboratory, 1 Cyclotron Rd., Berkeley, CA 94720, USA}
\author{D.Z. Osterman} \affiliation{University of Massachusetts, Amherst Center for Fundamental Interactions and Department of Physics, Amherst, MA 01003-9337 USA}
\author{P.K. Patel} \thanks{Corresponding author: \href{mailto:pratyush06853223015@gmail.com}{pratyush06853223015@gmail.com}} \affiliation{University of Massachusetts, Amherst Center for Fundamental Interactions and Department of Physics, Amherst, MA 01003-9337 USA}
\author{B. Penning} \affiliation{Physics Institute, Univ. of Z\"urich, Winterthurerst. 190, 8057 Z\"urich CH}
\author{H. Pinckney} \affiliation{University of Massachusetts, Amherst Center for Fundamental Interactions and Department of Physics, Amherst, MA 01003-9337 USA}
\author{M. Platt} \affiliation{Texas A\&M University, Department of Physics and Astronomy, College Station, TX 77843-4242, USA}
\author{Y. Qi} \affiliation{Department of Mechanical Engineering, FAMU-FSU College of Engineering, Florida State University, Tallahassee, FL 32310, USA} \affiliation{National High Magnetic Field Laboratory, Tallahassee, FL 32310, USA}
\author{M. Reed} \affiliation{University of California Berkeley, Department of Physics, Berkeley, CA 94720, USA}
\author{G.R.C Rischbieter} \affiliation{Physics Institute, Univ. of Z\"urich, Winterthurerst. 190, 8057 Z\"urich CH}\affiliation{University of Michigan, Randall Laboratory of Physics, Ann Arbor, MI 48109-1040, USA}
\author{R.K. Romani}\affiliation{University of California Berkeley, Department of Physics, Berkeley, CA 94720, USA}
\author{P. Sorensen} \affiliation{Lawrence Berkeley National Laboratory, 1 Cyclotron Rd., Berkeley, CA 94720, USA}
\author{V. Velan} \affiliation{Lawrence Berkeley National Laboratory, 1 Cyclotron Rd., Berkeley, CA 94720, USA}
\author{G. Wang} \affiliation{Argonne National Laboratory, 9700 S Cass Ave, Lemont, IL 60439, USA}
\author{Y. Wang} \affiliation{University of California Berkeley, Department of Physics, Berkeley, CA 94720, USA}\affiliation{Lawrence Berkeley National Laboratory, 1 Cyclotron Rd., Berkeley, CA 94720, USA}
\author{S.L. Watkins} \affiliation{University of California Berkeley, Department of Physics, Berkeley, CA 94720, USA}
\author{M.R. Williams} \affiliation{Lawrence Berkeley National Laboratory, 1 Cyclotron Rd., Berkeley, CA 94720, USA}

\collaboration{TESSERACT Collaboration}

\date{\today}

\begin{abstract}
Abstract: We describe design studies for a pulsed quasi-monoenergetic 2-keV neutron source for calibration of sub-keV nuclear recoils.  Such a calibration is required for detectors sensitive to sub-GeV dark matter and also the coherent elastic scattering of reactor neutrinos. In our design, neutrons from a commercial deuterium-tritium generator are moderated to the keV scale and subsequently filtered to achieve a monoenergetic spectrum utilizing the antiresonance feature, characterized by a drop in neutron transmission spectra at 2-keV for scandium. The analyses
 and conclusions presented are primarily based on simulations, as this study represents a critical first step toward the final goal. We address challenges such as unmoderated high-energy neutrons and gamma backgrounds from neutron capture in the moderator materials. Through optimization of the moderator+filter and shielding geometry, we identify a configuration that achieves the target neutron flux at 2~keV while maintaining subdominant rates of background interactions in simulation. Lastly, we outline a future pathway to lower-energy (few eV scale) calibrations using time-of-flight and sub-keV neutrons.

\end{abstract}

\maketitle

\section{Introduction} \label{sec:intro}
Several efforts are underway to search for dark matter (DM) down to the $\sim$keV mass scale, below the long-standing $>$~GeV focus of the field~\cite{PhysRevD.100.092007,PhysRevD.95.082002,anthonypetersen2023applyingsuperfluidheliumlight,fink2024superconductingquasiparticleamplifyingtransmonqubitbased,Griffin_2021,kim2024athermalphononcollectionefficiency}.  Several of these lower-mass efforts sensitive to DM elastically scattering with with nuclei, meaning calibration methods must be developed which produce low-energy nuclear recoils (NRs) in order to characterize the resulting material excitations~\cite{ricochetcollaboration2021ricochetprogressstatus,thenucleuscollaboration2022exploringcoherentelasticneutrinonucleus,adamski2024detectioncoherentelasticneutrinonucleus,aguilararevalo2024searchescenunsphysicsstandard, linehan2024listeningnewphysicsquantum, Ramanathan_2026}.  Similar detector technologies are also being developed to observe NRs induced by reactor neutrinos.  Both experimental efforts target NR energies in the sub-keV range, requiring neutrons at roughly the keV scale. The development of low-energy neutron sources will also be benefical to the Migdal effect ~\cite{PhysRevLett.127.081805}  search community using low-energy neutron scattering~\cite{PhysRevD.107.L041303,PhysRevD.110.092012,araujoMIGDALExperimentMeasuring2023}.

\begin{figure*}
    \centering
    \includegraphics[width=1\textwidth]
    {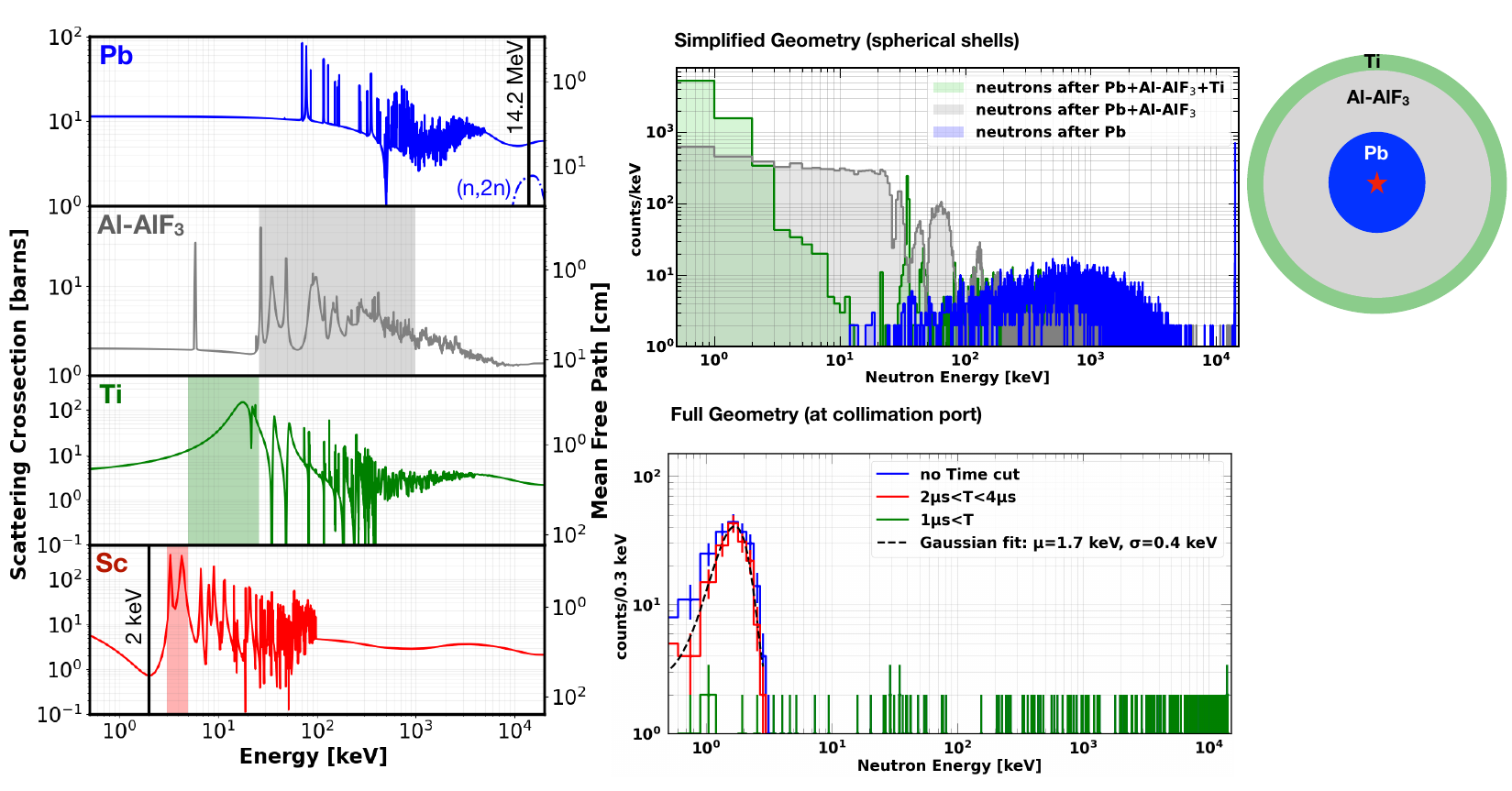}
    \caption{\emph{Left}: Neutron scattering cross sections and corresponding mean free paths for the materials used to produce the 2~keV neutron beam. These cross sections and mean free paths are evaluated using Ref.~\cite{Pritychenko2008Sigma, brownENDFBVIII08th2018,Liou1978ScTotalXS,Joki1961TiTotalXS}. The color-shaded regions denote the energy ranges over which each material is most effective in the moderation chain. \emph{Upper Right}: Evolution of the neutron spectrum after successive moderation stages. The blue spectrum corresponds to neutrons after the Pb converter, the grey spectrum to neutrons after Pb $+$ Al-AlF$_3$, and the green spectrum to neutrons after Pb $+$ Al-AlF$_3$ $+$ Ti. For this study, the materials were placed in a simplified spherical geometry surrounding a central point source of 14.1~MeV DT neutrons. \emph{Lower Right}: Neutron spectrum after transmission through the full moderator assembly and Sc filter. For this plot, the full detector geometry is used (see Fig.~\ref{fig:diffusion}), and the spectrum includes only neutrons passing through a 5~cm diameter surface at the collimation port. A timing selection of $2 < t < 4~\mu$s sharpens the 2~keV transmission peak (\textit{red curve}). In contrast, the spectrum with a timing selection of $t \leq 1~\mu$s is dominated by higher-energy neutrons (\textit{green curve}), demonstrating that timing information is largely retained throughout the moderation process and can be used to improve the spectral purity of the filtered neutron beam.}
    \label{fig:scattering_crossection}
\end{figure*}

As detector technologies push to eV scales, well-suited sources of monoenergetic neutrons become increasingly rare (energies are increasingly far from the `natural' MeV energy scale of nuclear processes).  Many techniques are under development for supplying low-energy neutrons; a non-exhaustive list of several complementary techniques includes:
\begin{enumerate}
    \item A convenient neutron source for this approach is a commercial deuterium-deuterium (DD) generator producing $\sim$2.45~MeV neutron~\cite{ArDD, LUXDD, HeDD}, a compact device producing pulsed neutrons in short, well-defined bursts rather than continuously, enabling coincidence tagging and time-of-flight measurements. Given the high neutron energy, only a small fraction of scatters fall in the very-small-angle sub-keV regime (see Fig.~\ref{fig:cartoon2}).  Lower-energy neutron beams can be produced at proton accelerators through reactions such as $^7$Li(p,n)$^7$Be or $^{51}$V(p,n)$^{51}$Cr.  The Triangle Universities Nuclear Laboratory (TUNL) and the Graaff Tandem accelerator facility at Universite de Montreal have supplied neutron beams with energies down to $\sim$50~keV for several recent NR calibrations~\cite{TUNL_Si, TUNL_NaI, TUNL_Xe, Montreal_pico}.
    \item The LUX and LZ experiments have supplemented a DD generator with a `reflector' material of low nuclear mass.  DD neutrons which (single-) scatter in the reflector show a strong correlation between reflection angle and energy, such that a selectively collimated population of scattered neutrons is both pulsed and mono-energetic at few-hundred-keV energies~\cite{DDreflector1,DDreflector2}.  A short DD pulse time and large separation distance between reflector and detector enables a calibration with minimal gamma backgrounds and a highly selective time of flight (TOF).
    \item Photoneutron sources can produce low-energy neutron fluxes, but with the cost of a non-pulsed constant flux and a high gamma background~\cite{photoneutronyields, cdms_photoneutron, Si_photoneutron, Si_photoneutron2}.  The TESSERACT collaboration has recently developed a novel improvement in which the gamma flux of a $^{124}$Sb--$^{9}$Be photoneutron source (E$_n$=24~keV) is shielded using Fe, a material with a low neutron scattering cross-section specifically at the 24~keV neutron energy~\cite{Biekert_2023}. Although photoneutron sources are not pulsed, their compact size can make them useful when the gamma rate is sufficiently low.
    \item When an excited nucleus emits a high-energy gamma ray, the nucleus recoils with a corresponding sub-keV energy.  Nuclei can be excited into these states through neutron capture, if the detector is placed in an environment of high thermal neutron flux. While the energies of this calibration can not be chosen (they are specific to the nucleus in question), the coincidence signal between NR and gamma emission can enable a clean tagging of the calibration recoils.  This strategy has been used by several groups to study the Ge NR response at 254~eV~\cite{jones1975,collar2021,kavner2024}.  It has been investigated for use in calibrating the liquid xenon response at several energies below 300~eV~\cite{gammaNR_LXe}.  The CRAB collaboration is developing a platform dedicated to this style of calibration for a variety of cryogenic detector targets~\cite{crab2, crab1}.
    \item At even lower energies, an MeV-scale gamma can scatter coherently with the target atom, resulting in a nuclear recoil~\cite{coherentgammas}.  The gamma scattering angle can be measured to reconstruct the NR energy.  This approach is promising, but has not yet been demonstrated.  
\end{enumerate}


This manuscript explores the design of a complementary neutron source through GEANT4~\cite{AGOSTINELLI2003250} simulation.  The strategy explored in this manuscript is complementary to the above approaches in that it is pulsed, supplies very low-energy neutrons (2~keV), and does not require bringing the detector to a large dedicated reactor or beam facility.  In Section~\ref{sec:Principle} we introduce the general working principles of the moderation and filtering of neutrons. In Section~\ref{sec:Design Studies of Filter} we discuss the geometry optimization procedure for the different moderators and filters used. The next section discusses the challenges of background rates both at the neutron backing detector and at the calibration target detector. In Section~\ref{sec:Calibrations} we discuss how this facility can be used to calibrate different target materials used in various sub-GeV dark matter search experiments. In the last section (Section \ref{sec:TOF_Calibrations}) we conclude by discussing a possible future evolution of these techniques to supply even lower-energy neutrons for future detectors.

\section{Neutron Moderation and Filtering to 2~keV}
\label{sec:Principle}

The neutron source is based on the neutron filter concept, together with effective neutron moderation. A neutron filter material has a special neutron-nucleus scattering cross section which exhibits a deep and narrow antiresonance at a characteristic energy. At this characteristic energy, the scattering cross section is uniquely small and the neutron mean free path is uniquely long, making the material \textit{transparent} to neutrons of this characteristic energy. Some neutron filters require isotopic enrichment, making them generally impractical.  The most practical neutron filter materials (in which a single isotope dominates in abundance) include silicon (149~keV)~\cite{tsangVersatileNeutronBeam1976}, sulfur (73~keV)~\cite{musolinoEvaluationIronfilteredEpithermal1991}, vanadium (59~keV)~\cite{JOSHI20146}, argon (57~keV~\cite{Andringa_2023}), iron (24~keV)~\cite{JOSHI20146}, and scandium (2~keV)~\cite{aminfarKeVScandiumFiltered1981}. These antiresonances are well understood using the Breit Wigner formalism~\cite{lynn1968theory}.  Implementing filters based on these materials is a long-established practice~\cite{filters1977, kyivfilters, Barbeau_2007}, though typically such implementations depend on an enormous initial neutron flux from a nuclear reactor.

The filter materials typically exhibit additional antiresonances at other energies.  In order to select only the lowest-energy antiresonance, the neutron flux incident on the filter material must first be tailored to the appropriate low-energy range.  The reactor-based filters mentioned above each use some tailored moderator materials before the filter material.

This manuscript describes an adaptation of the filter concept to a pulsed Deuterium-Tritium (DT) neutron generator as the initial neutron source.  As in the reactor case, we carefully moderate the initial high-energy flux to tailor it to the relevant energy range, then filter the neutrons using a filter material.  The commercial DT neutron generator has two major advantages over a reactor neutron source: it is compact and portable, and it is pulsed to supply a coincidence tag and reduce backgrounds to the calibration.  
 

\begin{figure}[h]
    \centering
    \includegraphics[width=0.5\textwidth]{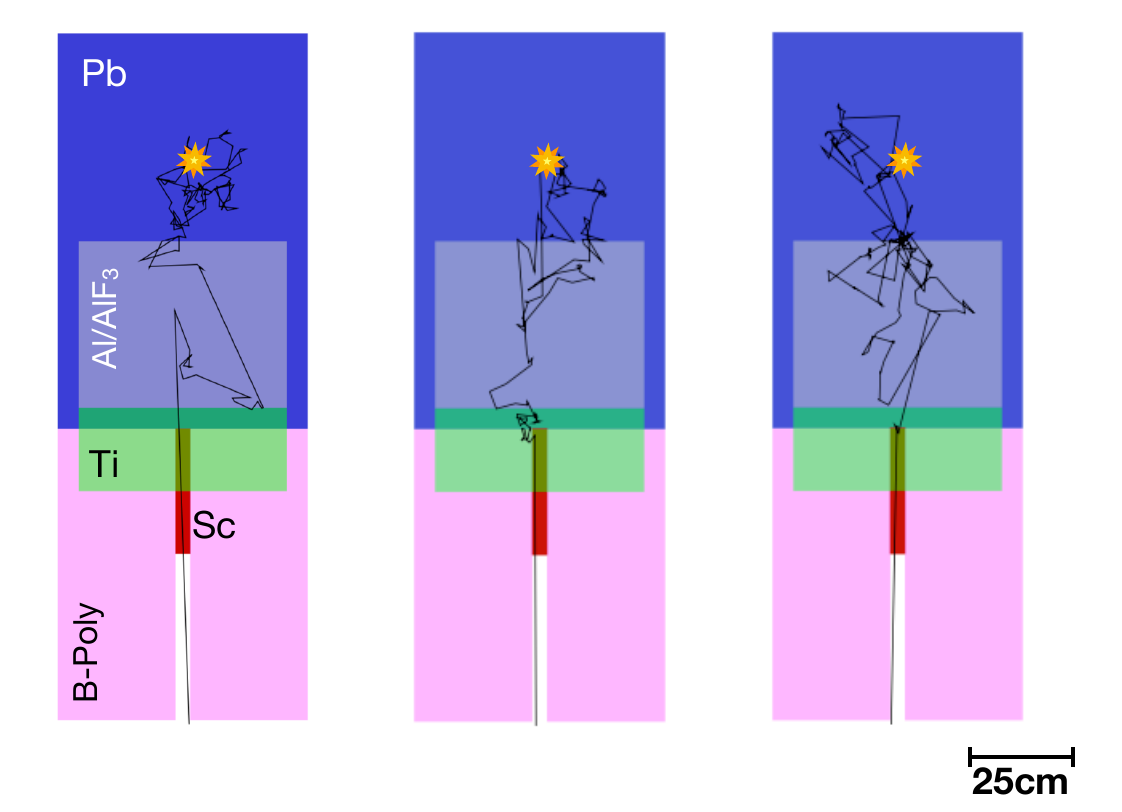}
    \caption{Three example neutrons from the 2~keV population of Fig.~\ref{fig:scattering_crossection}, showing the path of the 14.2~MeV neutron as it scatters many times in the moderating and filtering materials. The yellow-colored star represent the DT neutron source, where 14.2~MeV neutrons are bombarded uniformly in all directions.}
    \label{fig:diffusion}
\end{figure}

We describe in general terms the design goals and considerations for each important component used in the assembly, refer to Fig.~\ref{fig:scattering_crossection}. 

\textit{\textbf{Neutron source}}: The design starts with a compact pulsed neutron source. A pulsed Deuterium-Deuterium (DD) neutron generator may be the natural initial choice because the neutron energy is fairly low (2.45~MeV).  However, we have found that the moderating and filtering process in that case is neutron-starved.  The moderating and filtering process has an extremely low efficiency of $\mathcal{O}$($10^{-7}$), requiring the maximal initial neutron flux.  We have found that while the DT process is higher-energy (14.2~MeV), introducing additional human safety considerations and gamma backgrounds, the much larger flux of the DT process overcomes the low moderation/filtering efficiency and makes the strategy possible.  For this work, the specific DT neutron generator chosen was the VNIIA ING-10-20-120~\cite{VNIIA}, due to the short duration of the neutron pulses (0.8~$\mu$s) and the compact geometry (34~mm diameter).  The generator produces 10$^7$ neutrons per pulse.

\textit{\textbf{Lead}}: The initial moderation step in Pb is intended to reduce most neutron energies to the sub-MeV range. While the large nuclear mass of Pb means it is an inefficient elastic moderator, its large (n,2n) cross section (threshold: $\sim$10~MeV) allows not only for efficient moderation but also a boost in the total neutron flux available for subsequent steps~\cite{Fischer1989PbMultiplication, Nobre2026ENDFB81}. In order to minimize the geometric (1/r$^2$) loss in neutron flux, the process of producing neutrons with an energy of less than 1 MeV must be completed within a few tens of cm. The DT neutron generator is small in diameter and the Pb is closely-packed around it.  Together, the DT neutron generator and the surrounding Pb can be considered a pulsed source of $\sim$1~MeV neutrons.  (See Fig.~\ref{fig:scattering_crossection}, upper right.)

\textit{\textbf{Al-AlF$_3$ mixture}}: The next goal is to moderate the neutrons to the $\mathcal{O}$(10~keV) scale as efficiently as possible (in a short length, minimizing 1/r$^2$) through elastic scattering. This problem has been studied for the goal of delivering neutrons in boron neutron capture therapy (BNCT), where mixtures of aluminum and fluorine have been developed and characterized~\cite{bnct1, bnct2}. These two elements contain cross section resonances at complementary energies, giving a mixture of Al and F an unusually high neutron scattering cross section over a broad energy range (from $\sim$30~keV and above) and a sudden change to a low cross section below $\sim$30~keV.  Emulating this existing BNCT literature, we mixed Al and ALF$_3$ powders and then employed cold isostatic press (CIP) and hot isostatic press (HIP) methods to create a solid machinable material with a density of 2.5g/cm$^3$.  
    
\textit{\textbf{Titanium}}: The final moderation step fills in a \textit{gap} between the last moderation energy of the Al-AlF$_3$ (30~keV) and the goal neutron energy of 2~keV.  Titanium is a practical solution, thanks to its broad neutron resonance peaking just below the Al-AlF$_3$ range~\cite{Joki1961TiTotalXS,Schroder1985FilteredBeam}.  The combined effect of the Al-AlF$_3$ and the Ti is to moderate the majority of neutrons to $<$~10~keV, as seen in the lower-left plot of Fig.~\ref{fig:scattering_crossection}.
    
\textit{\textbf{Scandium}}: After the moderation steps result in a broad spectrum with $\mathcal{O}$(10~keV) maximum, that flux is passed to the filter material: Sc.  Neutrons of 2~keV energy exhibit a mean free path of 30~cm in Sc, estimated using the total neutron scattering cross-section data from ENDF/B-VIII.0 libraries~\cite{Liou1978ScTotalXS,brownENDFBVIII08th2018}.  Neutrons at very slightly higher energies have a dramatically shorter mean free path and will necessarily scatter within the Sc.  The antiresonance has a softer edge on the low-energy side, but still sharp enough to result in a quasi-monoenergetic spectrum (see Fig.~\ref{fig:scattering_crossection}, lower right).

\section{Simulation Studies: Optimization of Moderator and Filter Geometry}
\label{sec:Design Studies of Filter}

The geometry of the shielding components was optimized through detailed simulation studies performed using GEANT4 with a custom physics list. The
Shielding physics list~\cite{G4_physics} was selected to model neutron interactions from the ther-
mal regime up to 20 MeV. For thermal neutrons, the G4NeutronThermalScattering
process was enabled with an explicit thermal treatment of hydrogen, as this configuration provides improved accuracy in the low-energy regime relevant to the present
study. A transition energy of 4 eV was adopted, serving as both the upper limit
for thermal scattering and the lower limit for elastic scattering. This choice was
motivated by benchmark comparisons showing that, at thermal energies, GEANT4
results obtained with this configuration are in good agreement with MCNP simulations~\cite{goorleyInitialMCNP6Release2012}, with differences at the level of a few percent~\cite{VANDERENDE201640}. No additional systematic
comparison between GEANT4 and MCNP was performed within the scope of this design study beyond these established benchmarks. The first design goal was to maximize the flux and spectral purity of the 2~keV neutron flux leaving the Sc and the downstream collimator. {We define the neutron spectral purity as the fraction of outgoing neutrons in the energy window of 1-3 keV, since the transmitted neutron spectrum is well described by a Gaussian distribution with a mean energy of 1.7 keV and a standard deviation of 0.4 keV, corresponding to a full width at half maximum (FWHM) of approximately 0.94 keV (see the lower-right panel of Fig.~\ref{fig:scattering_crossection}). The 1-3~keV interval used in the definition of neutron spectral purity was chosen to encompass the majority of the transmitted neutron population around the 2~keV scandium transmission feature. Given the fitted Gaussian parameters, this interval corresponds approximately to the range $\mu \pm 1.6\sigma$, containing nearly 90\% of the transmitted neutrons. The bounds are therefore motivated by the observed width of the transmitted spectrum rather than being arbitrary. We expect modest variations of the window boundaries to change the absolute value of the reported spectral purity, but not the qualitative conclusions of the optimization study or the relative ranking of the moderator configurations considered. For purposes of this purity optimization step, we restrict the neutrons we consider to those passing through a small surface near the collimator port (5~cm diameter). Due to the large space of possible geometric considerations, we limited the geometric optimization to a variation of only a small number of geometric parameters. Scandium is an expensive material ($\sim$\$9,000 USD  per kg), hence the length of the filter is set by the neutron mean free path at 2~keV while the diameter (and volume) are set by material cost. A 99.5\% purity scandium rod, weighing 1~kg and measuring 3.5~cm in diameter and 
30 cm in length, was obtained. Prior to commencing the geometric optimization studies, several \emph{qualitative} geometry decisions were implemented, each of which were observed to benefit the purity. Examples of these qualitative decisions include the Al-AlF$_3$ being bound by Pb and 
much of the Sc being bound by Ti (see Fig.~\ref{fig:diffusion}). The subsequent geometry-optimization studies simultaneously varied four parameters:
\begin{enumerate}
    \item{Length of Pb (tested at 10, 15, 20, 25~cm)} 
    \item{Length of Al-AlF$_3$ (tested at 30, 35, 50~cm)}
    \item{Width of Al-AlF$_3$ (tested at 30, 40, 50, 60~cm)}
    \item{Length of Ti (tested at 3, 5, 7~cm)}
\end{enumerate}

\begin{figure}[h] 
\centering
\includegraphics[width=0.5\textwidth]
{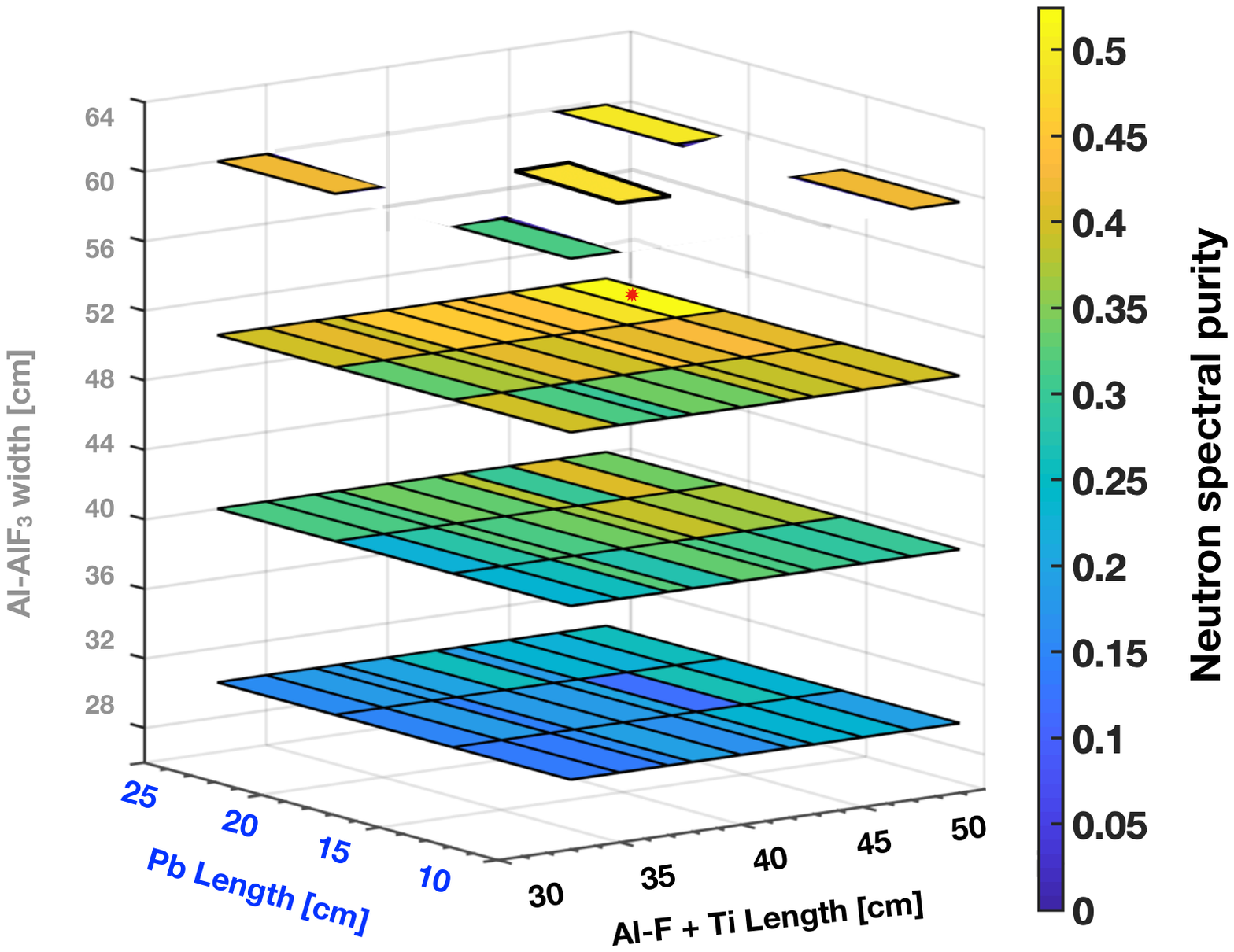}
\caption{A portion of the optimization space scanned via \textsc{GEANT4} simulation studies.  For each set of geometry values, the resulting neutron flux purity is characterized.  It is seen that increasing the quantity of all three moderator materials
will result in a higher neutron spectral purity, reaching
a maximum of $\sim$0.5. The width of the Al-AlF$_3$ moderator is seen to be of similar importance as the length in the beam direction. The red colored dot indicates the geometry that was chosen after this optimization process.}\label{fig:optimisation_geant4_studies}
\end{figure}

The width of the Pb, excluding the thickness that surrounds the Al-AlF$_3$, was kept similar to the width of the Al-AlF$_3$. Similarly, the width of the Ti was kept identical to the width of the Al-AlF$_3$. The thickness of Pb surrounding the Al-AlF$_3$ was not varied, as variation had no effect past a thickness of $\mathcal{O}$(1~cm). Fig. \ref{fig:optimisation_geant4_studies} illustrates the simulation results according to the 4-parameter scan described above.  For visual rendering, we collapse the space to 3 dimensions by summing two lengths to a single number:  the length of Al-AlF$_3$ plus the length of Ti.  Matching the intuitive expectation, we observe that the spectral purity increases as the dimension of the moderators increases. Similarly, we see a corresponding decrease in the flux, largely due to simple 1/r$^2$ behavior.  Some optimum must be chosen between the increase in purity and the decrease in flux.  Importantly, the purity reaches a maximum of approximately 50\%. The optimized geometry maximizes flux while maintaining 
50\% purity, resulting in the following dimensions: a Pb 
length of 25 cm, an Al-AlF$_3$ width of 50 cm (matching the Pb 
width), and a combined length of 47 cm for Al-AlF$_3$ and 
Ti, with Al-AlF$_3$ measuring 40 cm and Ti measuring 7 cm, summarized in Table~\ref{tab:moderation}. One key insight from the simulation effort was that the width of the moderating 
materials was as important as their length in the beam direction.  This is due to the diffusive nature of the neutron propagation as visualized in  Fig.~\ref{fig:diffusion}. 

\begin{table}[t]
\centering
\caption{Approximate neutron moderation chain from 14 MeV to below 10 keV. Neutrons are first down-converted in Pb via the $n,2n$ reaction, then moderated in Al-AlF$_3$ and Ti, while a Sc filter preferentially transmits neutrons near 2~keV. The mean free paths (MFPs) are representative order-of-magnitude estimates derived from the relevant energy-dependent cross sections from Fig.~\ref{fig:scattering_crossection}, left. The final column lists the final optimized material lengths obtained via GEANT4 optimization study.}
\vspace{0.2cm}
\label{tab:moderation}
\begin{tabular}{lccc}
\hline
Material & Energy Range & MFP & Length Used \\
         &              & (cm) & (cm) \\
\hline
Pb ($n,2n$) & 14~MeV $\rightarrow$ $<1$~MeV & $\sim15$ & 25 \\
Al--AlF$_3$ & $<1$~MeV $\rightarrow$ $<30$~keV & $\sim5$ & 40 \\
Ti & 30~keV $\rightarrow$ $<10$~keV & $\sim0.5$ & 7 \\
Sc filter & Filter 2~keV neutrons  & $\sim30$ & 30 \\
\hline
\end{tabular}
\end{table}

\section{Simulation Studies: Shielding and Backgrounds}
\label{sec:Shielding and Backgrounds}

\begin{figure}[h] 
    \centering
    \includegraphics[width=0.5\textwidth]
    {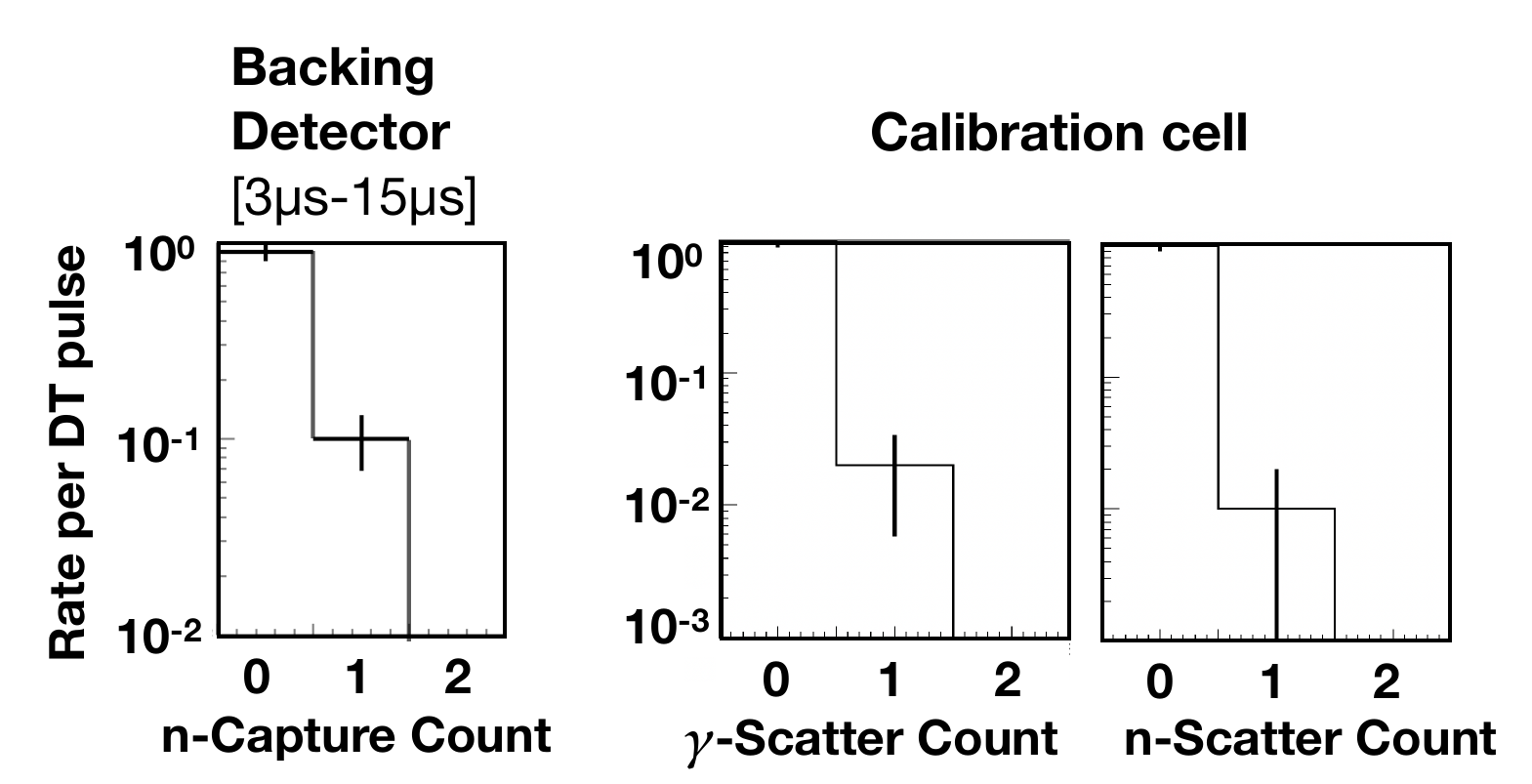}
    \caption{Number of counts per DT pulse of 10$^7$ neutrons.  These were the metrics of background suppression used to optimize the shielding geometry of Fig.~\ref{fig:renderings}.  In each category of background, the rate per DT pulse is $<$~0.1.}
    \label{fig:background_rates}
\end{figure}

\begin{figure*}
    \centering
    \includegraphics[width=1.00\textwidth]
    {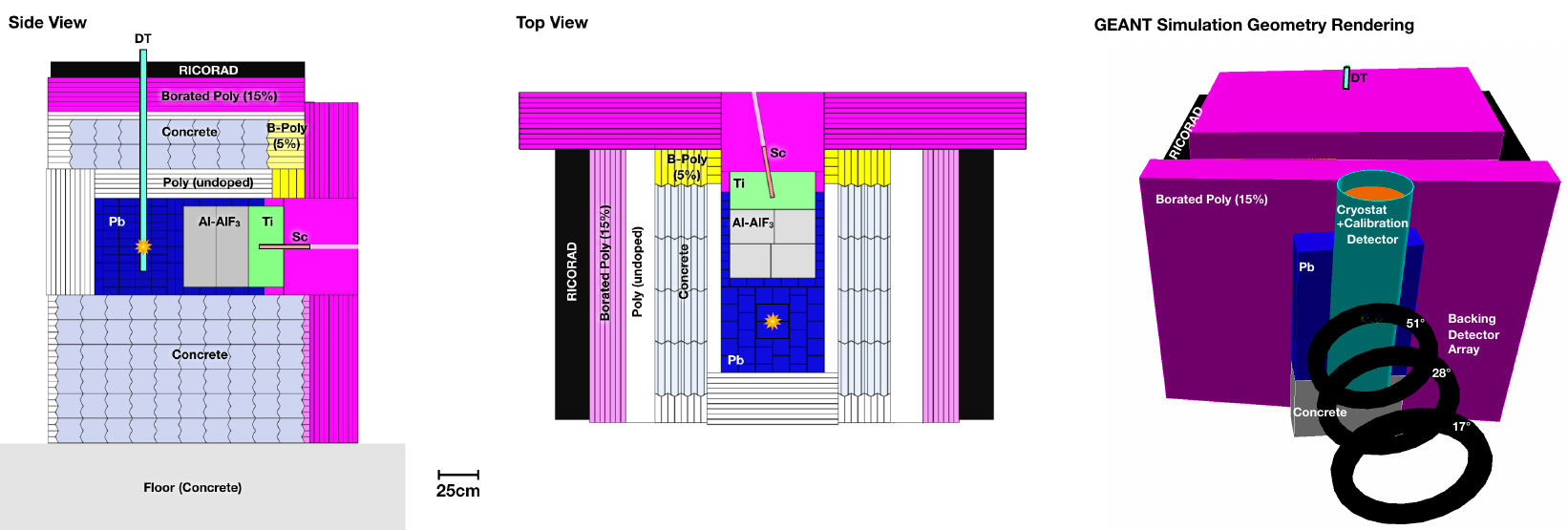}
    \caption{In order to suppress neutron and gamma backgrounds, significant quantities of shielding material were added surrounding the moderator and filter materials of Fig.~\ref{fig:diffusion}.  High-density concrete supports the moderator and filter assembly from below and is also used in a layer surrounding the filter.  Polyethylene (both boron-doped and undoped) is also relied on to moderate and capture neutrons.  Some amount of on-hand RICORAD supplements the borated poly.  In the Top view, one can see the 10-degree offset in the Scandium and collimation bore, which was seen to further suppress backgrounds.  The right view is a rendering of the simulation geometry used in the background and calibration studies (Sections IV and V), omitting only the building structures for clarity.}
    \label{fig:renderings}
\end{figure*}

Upon successful optimization of the elements defining the purity and flux of the 2~keV population, we separately consider the surrounding materials which determine the energy and flux of all other neutrons emitted from the filter assembly along with gammas from neutron capture processes on filter assembly materials.  Due to the high energy and high flux of the DT neutron generator, these \textit{background} neutrons and gammas pose a significant challenge.

In the strictest version of the calibration data selection, calibration events will satisfy a triple-coincidence between the DT neutron generator pulse time, the time of an energy deposit in the calibration detector, and the time of a neutron capture in the backing detector.  The timing of the background particles are well-correlated with the DT pulse time, however, so this triple coincidence is greatly weakened.  We must add significant quantities of material to the assembly to suppress the background neutron and gamma fluxes to the point that accidental coincidences are subdominant to true coincidences.

The quantity and quality of calibration data collected is limited in the following ways:

\begin{itemize}
    \item Due to the low efficiency of the moderation and filtering, 
and the finite operational lifetime of the generator before it 
needs to be replaced or refurbished, calibration statistics will 
be limited, making it essential to maximize calibration 
efficiency.
    
    \item Given the low neutron energy, the backing detector (for measuring neutron scattering angle) must be based on neutron capture rather than neutron scatter.  In order to maximize the calibration efficiency, the backing detector must also be of large solid angle and large capture efficiency. In previous work~\cite{BIEKERT2022166981} we have designed and tested a prototype of such a detector. Because the capture process follows an initial moderation step within the backing detector materials, the capture process is inherently slow, widening the coincidence window and increasing the rate of accidental coincidence. The large-area detectors of~\cite{BIEKERT2022166981} demonstrate a 2~keV neutron capture efficiency of 27\% and a mean neutron capture time of 17~$\mu$s.
    
    \item We must plan for calibration detectors based on phonon or heat detection, meaning the calibration detector will be slow and will also lead to a lengthening of the coincidence window.
    
\end{itemize}


 To quantify the expected beam-related coincident gamma and neutron background rates, a simulation was performed with the following laboratory settings:
\begin{itemize}
    \item{The DT generator and moderator+filter assembly (as illustrated in Fig.~\ref{fig:scattering_crossection} upper left) together with the surrounding shielding castle design under study}.
    \item{A calibration detector (a $^4$He target was assumed) and its associated dilution refrigerator infrastructure}.
    \item{An array of three backing detector rings of the type described in ~\cite{BIEKERT2022166981}}.
    \item{The surrounding building structures (walls, floor, ceiling, etc.)}.
\end{itemize}
To ensure a negligible rate of false pairings (between the calibration detector and the backing detector), a design goal was defined such that at most 10\% of DT pulses result in a neutron capture within the backing detector array within a specific time window of 3-15~$\mu$s following the initiation of the 1~$\mu$s DT pulse. This specific time window was chosen based on the simulation, as the majority of 
true coincidences show neutron capture in the backing detector array within the 3-15~$\mu$s window. 

We have investigated a variety of shielding materials, including pure water, a concentrated aqueous boron solution~\cite{TSUYUMOTO200720}, high density concrete, high-density polyethylene (HDPE), and boron-doped polyethylene. While simulations showed that water-based shielding could succeed at the design goals, practical considerations (e.g., the immersion of the DT neutron generator in water) drove the design towards solid-based options.  Once focusing on high-density concrete, HDPE, and boron-doped HDPE, it was found that multiple alternating layers of HDPE and high-density concrete provided superior background reduction compared to simpler geometries. The high-density concrete used in the setup has a density of 3.2 g/cm$^3$. The increased 
density compared to regular concrete is achieved by mixing iron aggregates into the 
grout/mortar. This particular batch contains approximately 40\% iron aggregate by weight. Undoped HDPE was used to moderate neutrons within the structure, and boron-doped polyethylene (combinations of 5\% and 15\% boron-doped polyethylene were used) was limited to the outer surface (where neutron energies should be low and capture efficiency should be high).  The thickness of the HDPE, concrete, and borated polyethylene were varied in simulation until the 10\% neutron capture probability design goal was reached as shown in Fig.~\ref{fig:background_rates}. Ambient accidental coincidences are expected to be negligible. For a representative backing-detector background rate of 10~Hz~\cite{BIEKERT2022166981} and a coincidence window of 10~$\mu$s, the probability of an accidental coincidence is approximately
$P_{\rm acc} \simeq R\Delta t = 10^{-4}$ per trigger window.
This is approximately three orders of magnitude smaller than
the expected neutron-induced coincidence probability
($\sim 10^{-1}$ per DT pulse, Fig.~\ref{fig:background_rates}); implying that accidental coincidences contribute only about 0.1\% of the observed coincidence events.

On the two side walls, the boron-doped polyethylene was supplemented with a layer of boron-doped polyvinyl chloride (RICORAD)~\cite{gaoLithium6FilterFission} because it was available on-hand.  As the boron concentration was uncertain in this material, the boron concentration was conservatively assumed to be 0\% in our studies. It was noticed that by including a purposeful misalignment of 10 degrees between the scandium and the rest of the structure, the background rates were reduced by a small but statistically measurable amount.  This 10 degree angle is illustrated in the center panel of Fig.~\ref{fig:renderings}.

Anticipating that in real construction incidental gaps may exist between materials, the alignment and shape of interfaces was considered in order to obstruct potential neutron leakage.  In particular, the concrete was purchased with a corrugated interlocking shape, and the polyethylene was cut to match the interlocking shape. To diminish the gamma rates at the calibration cell, an additional layer of lead, as illustrated in the central panel of Fig.~\ref{fig:renderings}, was positioned between the neutron-source shield and the calibration cell. The lead thickness was optimized to ensure that only a small percentage of DT pulses exhibit gamma scatter in the calibration cell, as illustrated in the right panel of Fig.~\ref{fig:background_rates}.

\section{Simulation Studies: Nuclear Recoil Calibration}
\label{sec:Calibrations}

Once the backgrounds were reduced below the design goals, the full geometry in the final form was used to simulate an example calibration dataset.

Each DT pulse emits 10$^7$ neutrons, and the goal calibration requires roughly 10$^7$ such pulses of the neutron generator.  A full simulation of 10$^{14}$ neutron emissions was not practical, so instead a two-stage \textit{boosting} procedure was applied, a strategy standard in shielding simulations (where a large initial flux in one region is reduced to a small final flux in another region).  In our case a hypothetical surface was constructed immediately surrounding the shielding and filtering materials.  Individual DT neutrons were generated, producing subsequent neutrons and gammas within these materials.  At this stage, the neutrons and gammas were simulated only until they passed through the surrounding hypothetical surface.  At that point the results of this \textit{stage 1} simulation were recorded.  Then these results served as the generator or seed for a \textit{stage 2} simulation, in which neutrons and gammas were emitted by the hypothetical surface.  In the first stage, 10$^{10}$ neutron emissions (equivalent to 10$^3$ DT pulses) were simulated.  In the second stage, those statistics were boosted by a factor of 10$^4$, meaning each gamma or neutron escaping the hypothetical surface was simulated 10$^4$ times.  In this way the complete statistics corresponding to 10$^{14}$ neutron emissions was achieved.

Understanding pileups and accidental coincidences requires grouping that simulation output into \textit{clusters} equivalent to individual DT pulses.  Using a lower-statistics full simulation, several Poisson mean expected counts per DT pulse were measured: the mean number of backing array detectors triggered, the mean number of neutron scatters in the calibration detector, and the mean number of gamma scatters in the calibration detector.  Clusters were assembled using these various mean counts and assuming Poisson fluctuations.

Finally, after clustering the energy deposits and neutron captures, the results were interpreted as they would appear when data-taking.  If multiple energy deposits occurred in the calibration detector, they were summed to a single large deposit.  If multiple neutron captures occurred in a single backing detector, they were treated as a single backing detector trigger.  A \textit{tagged} sample was constructed using events for which some energy was deposited in the calibration detector and some number of neutron captures appeared in one and only one backing detector.


We consider two data-taking approaches with different types of triggering or data selection:
\begin{enumerate}
    \item In an \textit{untagged} mode, the backing detector is not used.  We consider only the energy deposited in the calibration detector and the time of that energy deposit with respect to the DT pulse. (See Fig.~\ref{fig:cartoon2}, bottom panel.)
    \item In a \textit{tagged} mode, the backing detector is used to determine the scattering angle of the neutron.  As previously stated, we chose a backing detector neutron capture time of 3-15~$\mu$s. (See Fig.~\ref{fig:cartoon2}, upper panels.)  
\end{enumerate}

The results of this full calibration simulation are shown in Fig.~\ref{fig:cartoon2}.  It can be seen that the backing detector array has successfully tagged the scattering angles of a sufficiently large number of the 2~keV neutron population.  We also observe that neutron backgrounds (either higher-energy neutrons or neutron pileups in the calibration detector or backing detector array) are strongly sub-dominant.  More important is the effect of gamma backgrounds, included in the middle panel.  The effect of gammas is typically to enlarge an energy deposit in the calibration detector that would otherwise have been a well-tagged neutron scatter.  So in this way, the gamma backgrounds do not shift the tagged calibration peaks, but instead lower the statistics within the peaks. This emphasizes the importance of a low gamma background for calibration, and explains the motivation for the earlier design goal that only 10\% of DT pulses produce a gamma energy deposit in the calibration detector.

We end by emphasizing the main result: monoenergetic neutron scattering peaks are achievable down to a recoil energy of 10~eV, even for a low-mass target atom such as $^4$He.  For other target materials these tagged energies will be shifted lower by a factor of M$_{^{4}\text{He}}$/M$_{\text{target}}$.  For example, the $\sim$25~eV calibration peak in $^4$He would be shifted in Ge to $\sim$1.3~eV.

\begin{figure}[h!] 
\centering
\includegraphics[width=0.45\textwidth]
{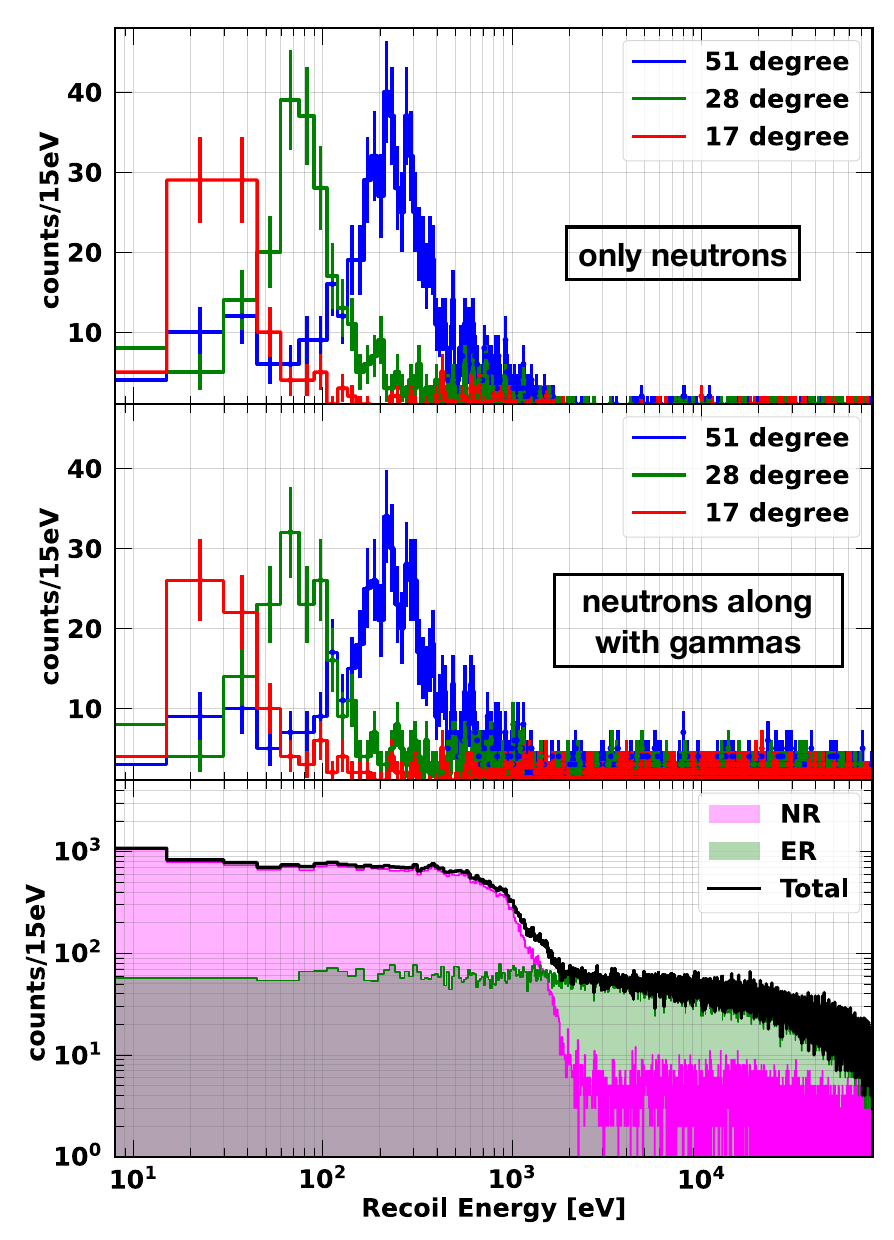}
\caption{\emph{Top:}  The simulated total energy deposited in a $^4$He calibration 
detector in \textit{tagged} data-taking mode, which includes energy 
deposits in the array of three backing detectors.  The observed peaks represent a successful tagging of the recoil angle of the quasi-monoenergetic 2~keV flux.  While there are also higher-energy backgrounds, that spectrum is roughly flat and extends to quite high energies, meaning it does not interfere with the characterization of the tagged peaks. In this top panel, only neutrons are considered; gamma backgrounds (largely from neutron capture) have been ignored. \emph{Middle:} The same as the top plot, but now including the effect of coincident gamma depositions in the calibration detector. \emph{Bottom:} In the \textit{untagged} mode, when the backing detector array is not used, we see the full spectrum of NR and gamma scatters in the calibration detector.  The statistics are much greater in this mode, and the endpoint energy provides some useful energy calibration.  In all three panels, 10$^7$ pulses of the generator are assumed, each emitting 10$^7$ neutrons.}\label{fig:cartoon2}
\end{figure}

\section{A future variation:  lower energies and Time of Flight}
\label{sec:TOF_Calibrations}
The scandium antiresonance at 2~keV represents the lowest-energy practical monoenergetic neutron filter.  It is fortunate, then, that at lower energies, neutron time-of-flight (TOF) becomes sufficiently 
long even in a university lab setting. This allows for the use of a non-monoenergetic 
neutron source, with neutron energy determined on a neutron-by-neutron basis through 
timing. In this section we describe some preliminary studies of this future calibration architecture, employing non-monoenergetic neutrons of sub-keV energy.

In the non-relativistic limit, one can write a convenient expression for the neutron TOF~=~72.3~$L/\sqrt{E_n}$ where $E_n$ is the neutron energy in eV, $L$ is the travel distance in meters, and TOF is measured in $\mu$s.  To make this even more concrete, we can report TOF for a 2~m travel distance, a short distance convenient for any laboratory setting.  The TOF for a 10~eV neutron would be 46~$\mu$s, and while for a 1~eV neutron it is  145~$\mu$s.  These times are long in comparison to the other timescales involved:  the 0.8~$\mu$s DT pulse duration, the $\sim$1~$\mu$s delays incurred during moderation, the timing resolution of typical calibration detectors, and the $\mathcal{O}$(10~$\mu$s) neutron capture time of the backing detector array as in ~\cite{BIEKERT2022166981}. Note that the neutron capture time of the backing detector can be shortened by reducing its 
thickness, though this comes at the expense of neutron capture efficiency. In summary, assuming a source of neutrons in the $E_n\lesssim$ 100~eV range, the TOF and hence $E_n$ can be measured on an event-by-event basis and the neutron source no longer benefits from being monoenergetic.

Much of the moderating materials and geometry of the Sc filter can be repurposed for this TOF-based calibration.  The new design goal would be to instead supply a broad spectrum of neutrons specifically within the $E_n\lesssim$~100~eV range, with strong suppression of neutrons at higher energies.  We note that reference~\cite{areg} has already demonstrated the key idea of a pulsed generator paired with a moderator to produce a pulsed source of broad-spectrum neutrons.  In that work, polyethylene served as the moderating material. Given our sensitivity to neutron pileup, we require a stronger suppression of the higher-energy flux and a more narrow focus of the spectrum towards the $E_n\lesssim$~100~eV range.

Simulation studies were performed to study the feasibility of obtaining a useful $E_n\lesssim$~100~eV neutron spectrum.  For simplicity, in this work we used only a toy geometry of nested spherical layers of moderating materials. The initial materials are exactly as previously optimized for the 2~keV Sc filter:  the DT neutron generator is surrounded by Pb, followed by Al-AlF$_3$ and Ti. As shown in the lower left of Fig.~\ref{fig:scattering_crossection} , this sequence results in a neutron spectrum largely below 10~keV.  Further moderation must now be applied to remove the $E_n\gtrsim$~100~eV population.  We find that a final moderating layer of Mn can achieve the goal.  Mn has a neutron scattering cross-section that is relatively high in the relevant $E_n\gtrsim$~100~eV range and relatively low below~\cite{Schroder1985FilteredBeam}.  A manganese layer of 10~cm thickness results in a roughly flat neutron spectrum up to several hundred eV. In Fig.~\ref{fig:TOF} we show the final neutron spectrum, along with the result of a polyethylene moderator of same thickness for comparison.  We also show the result of an example calibration employing TOF.
The comprehensive design of an NR calibration facility that utilizes TOF will be thoroughly examined in forthcoming works.

\begin{figure}[h!] 
\centering
\includegraphics[width=0.45\textwidth]{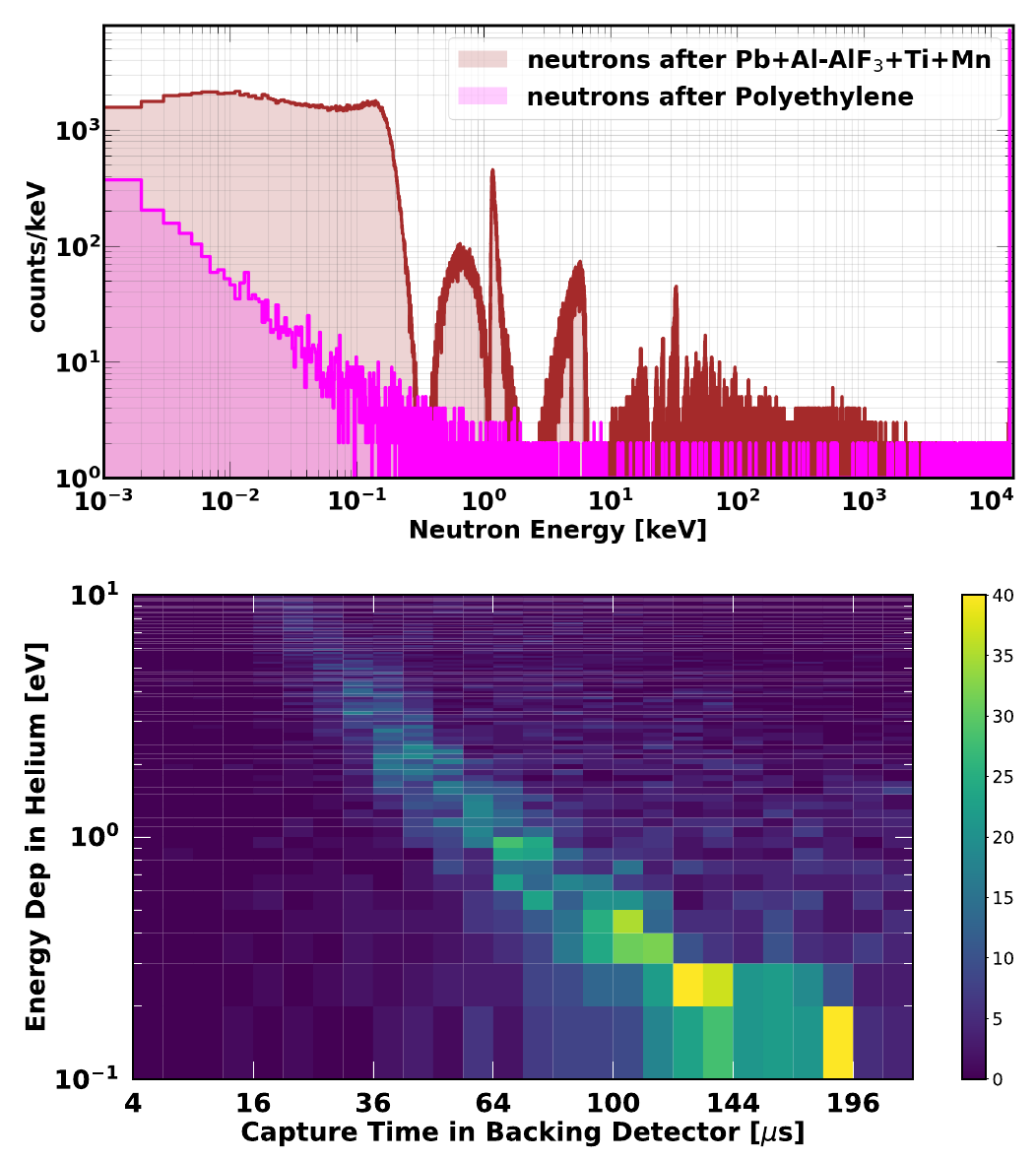}
\caption{\emph{Top:} The moderating effect of Pb + Al-AlF3 + Ti + Mn (brown) on neutrons is compared to the moderating effect of polyethylene (magenta). The study involved placing the materials in a simplified spherical shape around a central point-source of 14.1 MeV DT neutrons. \emph{Bottom:} Results of a basic simulation study conducted for TOF-calibration. In this study, an energy cut of 0.1~eV is applied to the energy deposits.}\label{fig:TOF}
\end{figure}

\section{Conclusion}
\label{sec:Conclusion}

The design of a keV neutron source for calibration of sub-keV NR in various cryogenic targets
has been successfully demonstrated. The primary objective of this study was to provide a comprehensive description of an advanced technique for the moderation and filtration of neutrons at the keV scale. 
The aforementioned research has demonstrated that the presence of a neutron source 
with an energy scale in the keV range alone is not sufficient to achieve a high rate of 
tagged sub-keV nuclear recoils. It is imperative to implement appropriate shielding measures in order to prevent the accumulation of unwanted signals in these detectors that rely on slow phonons. The investigation of the formation of the Al-AlF$_3$ combination has also encompassed an examination of its practical implications. Both the low mass dark matter detectors and the coherent elastic neutrino-nucleus scattering (CE$\nu$NS) detector group will derive advantages from the establishment of this nuclear recoil calibration facility. This facility will serve as a platform for conducting sub-keV nuclear recoil calibration experiments.   Enhancements to the DT neutron generator and its accompanying backing neutron detector have the potential to decrease the detection threshold for nuclear recoils. 

\begin{acknowledgments}
This work was supported in part by DOE Grant DE-SC0019319, DE-SC0025523, DE-SC0010004, DE-SC0020374 and DOE Quantum Information Science Enabled Discovery (QuantISED) for High Energy Physics (KA2401032). This research was supported in part through computational resources and services provided by Advanced Research Computing at the University of Michigan, Ann Arbor. This work was also completed in part with resources provided by the University of Massachusetts' Green High Performance Computing Cluster (GHPCC). We also thank Areg Danagoulian, Andreas Biekert and Alan Robinson for helpful discussions. Y.Q. and W.G. acknowledge the support from the National High Magnetic Field Laboratory at Florida State University, which is supported by the National Science Foundation Cooperative Agreement No. DMR-2128556 and the state of Florida.

\end{acknowledgments}

\bibliographystyle{apsrev4-1}
\bibliography{KeVScale_Neutron_Source_NIMA.bib}

\end{document}